\documentclass{optica-article}

\journal{oe}

\begin{document}

\title{Inverse saturable absorption mechanism and design optimization of mode-locked fiber lasers with a nonlinear amplifying loop mirror}

\author{XIANG ZHANG,\authormark{1,2,3} YONG SHEN,\authormark{1,2,3} XIAOKANG TANG,\authormark{1,2} QU LIU,\authormark{1,2} and HONGXIN ZOU\authormark{1,2,*}}

\address{\authormark{1}Institute for Quantum Science and Technology, College of Science, National University of Defense Technology, Changsha 410073, China\\
\authormark{2}Hunan Key Laboratory of Mechanism and Technology of Quantum Information, Changsha 410073, China\\
\authormark{3}The authors contributed equally to this work.}

\email{\authormark{*}hxzou@nudt.edu.cn} 



\begin{abstract*}
From the point of view of the differential phase delay experienced by the two counterpropagating optical fields, the self-starting of the mode-locked fiber laser with a nonlinear amplifying loop mirror (NALM) is theoretically studied. Although it is generally believed that NALM shows a saturable absorption effect on both continuous wave (CW) light and pulses, we find a counter-intuitive fact that cross-phase modulation (XPM) leads to opposite signs of differential nonlinear phase shifts (NPSs) in these two cases, resulting in inverse saturable absorption (ISA) during pulse formation process.
The ISA is not helpful for the self-starting of laser mode-locking and can be alleviated by introducing a non-reciprocal phase shifter into the fiber loop. 
In addition, we analyze the influences of gain-fiber position, splitting ratio, and optical attenuator in the fiber loop, on the differential NPS and self-starting process. 
These results are helpful for optimizing the design of NALM and lowering the self-starting threshold of the high-repetition-rate mode-locked fiber laser.
\end{abstract*}

\section{Introduction}
Due to their low cost, high stability, and compactness, mode-locked fiber lasers have been widely studied and enabled various applications in recent years. The NALM\cite{Fermann1990}, as a combination of an optical switch nonlinear optical loop mirror (NOLM)\cite{Doran1988} and a fiber amplifier, is often used as the artificial saturable absorber to achieve mode-locking in fiber lasers\cite{Duling1991a,Fermann1991,Richardson1991}. In addition, a loss is introduced into the NOLM to make a new asymmetric fiber loop referred to as nonlinear absorbing loop mirror (NAbLM)\cite{Bulushev1991,Seong2007,Zhao2020d}. The NOLM and NALM are originally proposed as optical switches, and many optimization methods have been proposed to improve the asymmetry of the fiber loops to facilitate the optical switching function, such as adding absorption loss\cite{Bogoni2004,Pitois2005,Seong2007} or output loss\cite{Bulushev1991,Ai2011}, making loop dispersion imbalance\cite{Steele1993,Wong1997a,Matsumoto1998,Yongkui2003}, using unbalance splitter\cite{Mortimore1988,Steele1996,Shah2001,Meissner2005}, incorporation of a nonlinear element\cite{Kane1994,Sponsel2007,Hierold2011,Zhang2021c}, introducing a nonreciprocal phase bias\cite{HongLin1994,Hierold2011,Zhang2021c}, and using fiber birefringence\cite{Doran1988,Moores1991,Finlayson1992,HeeYealRhy2000,Pottiez2004a}. In the last ten years, these methods have also been used to improve the performance of mode-locked fiber lasers with optical loop mirror.

As far as we know, the state of the art mode-locked erbium-doped fiber lasers with a NALM still have a low repetition rate. For the non-polarization-maintaining scheme, the highest is 257 MHz\cite{WanliGao2018}. For the polarization-maintaining scheme, the highest is 250 MHz\cite{Hansel2017}. The main limiting factors are the low erbium-doping concentration of the gain fiber, the limited fiber loop length, and the resulting low gain. Besides, the optimization methods mentioned above are not always helpful for the self-starting of high repetition rate mode-locked lasers. It is necessary to find a more effective method from the saturable absorption mechanism of optical loop mirror.

For the optical-switch NALM and NOLM, many researchers have found that the XPM effect related to the energy distribution and encounter of lights in fiber loop affects the performance of these two switches\cite{Bogoni2004,Pitois2005,Jinno1992,Clausen1996} and even inverts the switch function\cite{Finlayson1992}. While for the mode-locked fiber laser with a NALM or NOLM, the self-starting of mode-locking is closely related to the differential NPS accumulation of two counterpropagating beams in the fiber loop. Thus, the XPM effect definitely makes a contribution to pulse formation and self-starting in the mode-locked fiber lasers, which has not been thoroughly studied yet.

In order to overcome the difficulties of increasing the repetition rate and analyze the influence of XPM on self-starting of mode-locking, we calculate the power distributions of two counterpropagating beams in the NALM and evaluate the effects of configuration parameters, such as gain-fiber position and loop loss, on the mode-locking self-starting. We find a difference between the accumulated differential NPSs for the CW light and the pulses in the fiber loop, which makes the NOLM or NALM show an ISA mechanism during the pulse formation. The ISA has been extensively studied in the real saturable absorber\cite{Keller2021,Malcuit1984,Rakov2001,Li2014,Cheng2015,Tian2018}, but not in artificial saturable absorber. The ISA in NOLM or NALM could be used to explain the experimental phenomena that the mode-locking of laser can be actively started by tapping fiber, fine-tuning light polarization, or other disturbances.

\section{Theoratical calculation of EDFA and NPS}
The beam propagation in the erbium doped fiber amplifier (EDFA) of a NALM is obtained by employing the Giles model\cite{Giles1991}. The Giles model describes an active fiber as a two-level system using four spectroscopic parameters, including the absorption spectrum $\alpha$, the gain spectrum $g^*$, the saturation parameter $\zeta$, and the linear loss $l$. The optical power of the $k$-th beam $P_k(z)$ at position $z$ and the ratio of ions in the excited state ${\bar{N}_2} {/} {\bar{N}_1}$ are described as
\begin{equation}
\frac{dP_k}{dz}=u_k\left( \alpha _k+g_{k}^{*} \right) \frac{\bar{N}_2}{\bar{N}_t}P_k\left( z \right) +2u_kg_{k}^{*}\frac{\bar{N}_2}{\bar{N}_t}h\nu _k\Delta\nu_k-u_k\left( \alpha _k+l_k \right) P_k,
\label{eq:dP_k}
\end{equation}
\begin{equation}
\frac{\bar{N}_2}{\bar{N}_t}=\frac{\Sigma _k\frac{P_k(z)\alpha _k}{h\nu _k\zeta}}{1+\Sigma _k\frac{P_k(z)(\alpha _k+g_{k}^{*})}{h\nu _k\zeta}},
\label{eq:N_2}
\end{equation}
where $h\nu _k$ is the photon energy, $\Delta \nu$ is the frequency bandwidth of the beam, $u_k = \pm 1$ denotes the propagation direction of the beam.

The nonlinear effects considered in our model include self-phase modulation (SPM) and XPM. SPM refers to the self-induced phase shift experienced by an optical field $E$ during its propagation in optical fibers, whose magnitude is
\begin{equation}
\phi _\mathrm{SPM}=\bar{n}_2k_0\int_0^L{\left| E\left( z \right) \right|^2}dz=\gamma \int_0^L{P}\left( z \right) dz,
\label{eq:phi_SPM}
\end{equation}
where $\bar{n}_2$ is the nonlinear-index coefficient, and $\gamma ={{2\pi n_2}{/}{\left( \lambda A_\mathrm{eff} \right)}}$ is the nonlinear coefficient. In our system, $\gamma _1$ of the passive fiber is 1.367 $\rm{W}^{-1}{/}\rm{km}$, and $\gamma _2$ of the erbium-doped fiber is 2.134 $\rm{W}^{-1}{/}\rm{km}$. XPM refers to the differential NPS of an optical field induced by another field with a different wavelength, propagation direction, or polarization. Since the beams in NALM have the same polarization and counter propagate, the XPM coefficient is equal to 2. Other nonlinear effects  except for SPM and XPM are not considered in the following calculations.
\begin{figure}[htbp]
	\centering\includegraphics[width=0.5\textwidth]{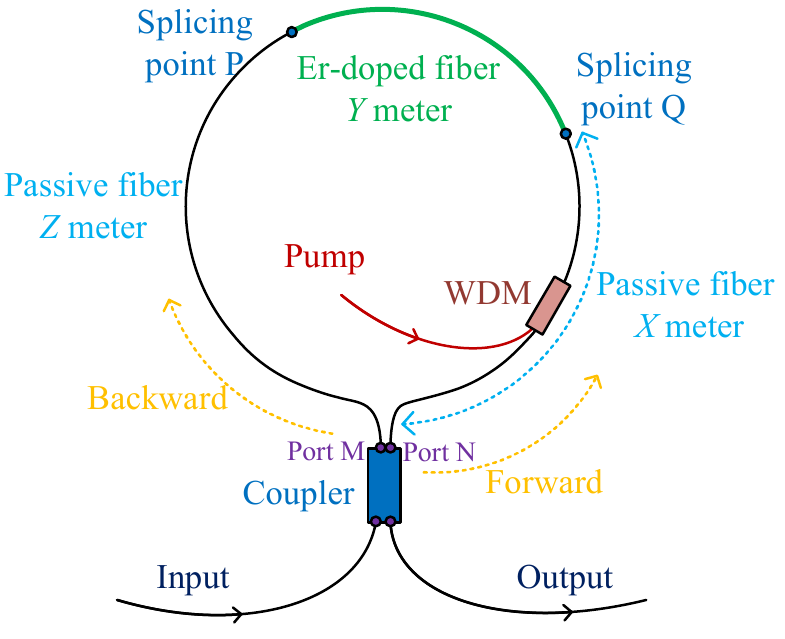}
	\caption{The configuration of NALM. The length of passive fiber from port N to splicing point Q is $X$ meter. The length of Er-doped fiber is $Y$ meter. The length of passive fiber from port M to splicing point P is $Z$ meter. The counterclockwise beam and the clockwise beam are named forward beam and backward beam, respectively. Their optical intensity ratio is $\rho:(1-\rho)$.}
	\label{fig1}
\end{figure}

The configuration of NALM is shown in Fig.~\ref{fig1}. Since the gain fiber takes up a large portion of the loop, the influence of optical power distribution in the gain fiber on the differential NPS accumulation needs to be considered. Before pulse formation, the light in the cavity is a continuous wave with low power. Thus, we regard the initial light as CW light with amplitude $A_0$. As such an optical signal enters the input port of the NALM, amplitudes of the forward and backward propagating fields are given by
\begin{equation}
	A_f(0)=\sqrt{\rho}A_0,\quad A_b(0)=i\sqrt{1-\rho}A_0.
	\label{eq:A_f}
\end{equation}
After one round trip, both fields acquire a linear phase shift and differential NPSs induced by SPM and XPM\cite{Agrawal2019}, and can be written as
\begin{equation}
A'_{f}=\sqrt{\rho}\sqrt{g_f}A_0\exp\mathrm{(}i\phi _0+i\phi _f),\quad
A'_{b}=i\sqrt{1-\rho}\sqrt{g_b}A_0\exp\mathrm{(}i\phi _0+i\phi _b),
\label{eq:A'_f}
\end{equation}
where $\phi _0$ is the linear phase shift, $g_f$ and $g_b$ are the gain factors of the forward and backward fields, respectively. Since the fiber is short, the fiber loss is neglected in the calculation.

As the two beams are output from the M and N ports, respectively, and counter propagate in fiber, the SPM phase shifts start to be generated. Then they meet in the gain fiber and indirectly interact with each other through the medium to generate the XPM phase shifts. Thus, the NPSs of the forward and the backward beams can be divided into four parts and expressed by SPM and XPM terms as
\begin{equation}
\begin{aligned}
	\phi _f=&\gamma _1\rho |A_0|^2X+\gamma _2\int_X^{\frac{X+Y+Z}{2}}{\left| A_f\left( z \right) \right|^2}dz+\gamma _2\int_{\frac{X+Y+Z}{2}}^{\frac{3X+Y+Z}{2}}{\left| A_f\left( z \right) \right|^2}+2\left| A_b\left( z \right) \right|^2dz\\
	&+\gamma _1\left[ \rho g_f+2\left( 1-\rho \right) \right] |A_0|^2Z,
\end{aligned}
\label{eq:phi_f}
\end{equation}

\begin{equation}
	\begin{aligned}
	\phi _b=&\gamma _1\left[ \left( 1-\rho \right) g_b+2\rho \right] |A_0|^2X+\gamma _2\int_X^{\frac{X+Y+Z}{2}}{2}\left| A_f\left( z \right) \right|^2+\left| A_b\left( z \right) \right|^2dz\\
	&+\gamma _2\int_{\frac{X+Y+Z}{2}}^{\frac{3X+Y+Z}{2}}{\left| A_b\left( z \right) \right|^2}dz+\gamma _1(1-\rho )|A_0|^2Z,
	\end{aligned}
\label{eq:phi_b}
\end{equation}
and the differential NPS is
\begin{equation}
	\begin{aligned}
		\phi _f-\phi _b=&\gamma _1|A_0|^2X\left( \rho g_b-g_b-\rho \right)-\gamma_2\int_X^{\frac{X+Y+Z}{2}}{\left| A_f\left( z \right) \right|^2}+\left| A_b\left( z \right) \right|^2dz\\
		&+\gamma _2\int_{\frac{X+Y+Z}{2}}^{\frac{3X+Y+Z}{2}}{\left| A_f\left( z \right) \right|^2}+\left| A_b\left( z \right) \right|^2dz+\gamma _1|A_0|^2Z\left( \rho g_f-\rho +1 \right),
	\end{aligned}
\label{eq:phi_f-phi_b}
\end{equation}
where we assume $X < (X+Y+Z)/2 < X+Y$, $z$ is the distance from the light wavefront to port M, $A_f (z)$ and $A_b (z)$ are the amplitudes of forward and backward light at $z$, respectively. Before self-starting of the mode-locked laser, the light in the cavity is CW, whose differential NPS can be obtained with Eq.~(\ref{eq:phi_f-phi_b}). Since only two counterpropagating pulses exhibit in the cavity when the laser modes are locked, the XPM phase shift is only generated where the pulses meet and can be neglected. Therefore, the differential NPS of the pulsed light is only related to SPM and can be expressed as
\begin{equation}
	\begin{aligned}
		\phi _f-\phi _b=&\gamma _1|A_0|^2\left[ \rho g_f+\rho -1+X\left( 1-g_b+\rho g_b-\rho g_f \right) \right] 
		\\
		&+\gamma _2\int_X^{\frac{3X+Y+Z}{2}}{|}A_f(z)|^2-|A_b(z)|^2dz.
	\end{aligned}
\label{eq:f-b_pulsed}
\end{equation}
When these two pulses meet at the fiber coupler, using Jones matrix, the reflected and transmitted fields can be given by
\begin{equation}
	\left( \begin{array}{c}
		A_t\\
		A_r\\
	\end{array} \right) =\left( \begin{matrix}
		\sqrt{\rho}&		i\sqrt{1-\rho}\\
		i\sqrt{1-\rho}&		\sqrt{\rho}\\
	\end{matrix} \right) \left( \begin{array}{c}
		A'_{f}\\
		A'_{b}\\
	\end{array} \right) .
\label{eq:At_Ar}
\end{equation}
Then the transmittance $T\equiv|A_t|^2/|A_0|^2$ of the NALM is
\begin{equation}
	T=\rho ^2g_f+(\rho -1)^2g_b+2\rho (\rho -1)\sqrt{g_fg_b}\cos\mathrm{(}\phi _f-\phi _b).
\label{eq:T}
\end{equation}
It is clear that, unlike the differential NPS and the gain factors, the $\phi_0$ term in Eq.~(\ref{eq:A'_f}) has no contribution to the transmittance. A similar transmittance expression can be obtained in a figure-9 configuration with a polarizing beam splitter (PBS) in place of the fiber coupler in Fig.~\ref{fig1}\cite{Duan2020}.

\section{Results and discussion}
\subsection{Inverse saturable absorption mechanism in NALM}
\begin{figure}[htbp]
	\centering\includegraphics[width=0.75\textwidth]{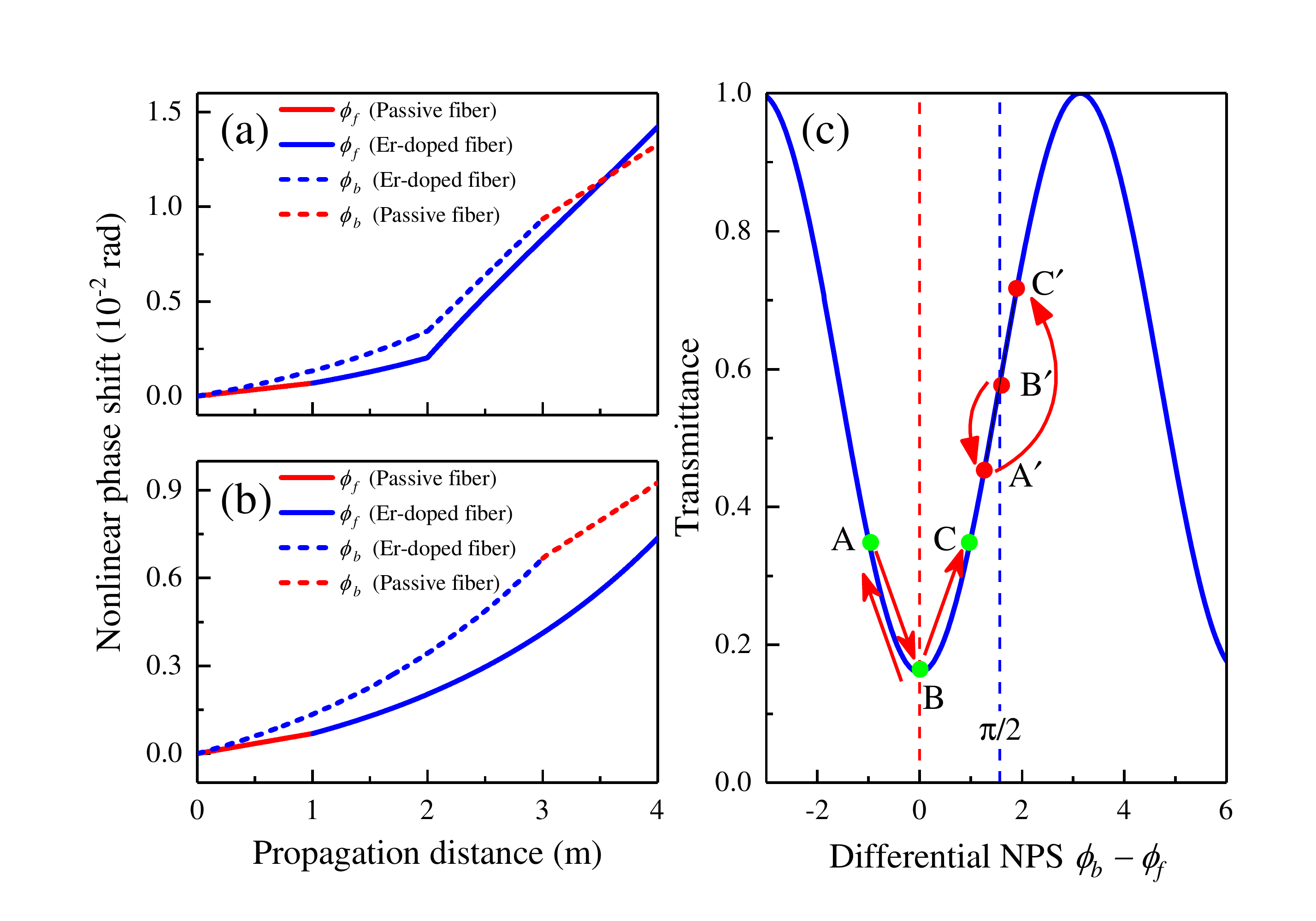}
	\caption{(a) NPS of the CW light in NALM. (b) NPS of the pulsed light in NALM. (c) The NOLM transmittance as functions of the differential NPS. For the NALM, the splitting ratio is 0.5, the passive fiber length $X = 1$ m, $Z = 0$ and the Er-doped fiber length $Y = 3$ m. Point B is at $\phi_f-\phi_b=0$, Point B$’$ is at $\phi_f-\phi_b=\pi /2$. Points A and A$'$ are differential NPS inflection points, and point A$'$ is not on the left side of point B. Points C and C$'$ are any points of the uphill curve, to the right of points B and B$’$, respectively.}
	\label{fig2}
\end{figure}

We assume that the peak power of the pulsed light is equal to the power of the CW light. Based on Eqs.~(\ref{eq:phi_f-phi_b}) and (\ref{eq:f-b_pulsed}), we can get the NPSs of the two beams in the NALM, which are shown in Figs.~\ref{fig2}(a) and \ref{fig2}(b), respectively. It is evident that the difference between phase delay $\phi_f$ and phase delay $\phi_b$ of CW light decreases to zero and becomes positive as a result of XPM after the two beams meet, whereas the differential NPS of pulsed light increases further. In the NALM mode-locked fiber laser, as the power of the light in the cavity increases, the light evolves from CW to pulses, and the differential-NPS-sign transition can occur during the self-starting of mode-locking.

We discuss the influence of this differential-NPS-sign transition on the saturable absorption effect of the NALM and self-starting of mode locking based on the transmittance curve of NOLM given in Fig.~\ref{fig2}(c). Here, we choose to discuss NOLM instead of NALM for convenience. In the process of optical signal changing from CW to pulses, point B at $\phi_f-\phi_b=0$ has two roles. One is the point where the pump is turned on, and the other is the point where the sign of the differential NPS changes. When the pump is turned on, the optical signal is continuous, and the round-trip differential NPS increases along the trace B $\rightarrow$ A, which is manifested as positive feedback with less loss for the stronger light and more loss for the weaker light. Therefore, it is the saturable absorption effect that favors the formation of pulses. When the light gradually evolves from CW to pulses, the differential NPS moves along  A $\rightarrow$ B $\rightarrow$ C, as depicted in Fig.~\ref{fig2}(c). As the peak-to-average power ratio of the pulse gradually increases, the SPM effect dominates over the XPM effect, resulting in a change in the sign of the differential NPS. In the A $\rightarrow$ B process, the fiber loop shows negative feedback, i.e., the stronger the light, the greater the loss. Therefore, the so-called ISA effect merges, which is not helpful to the formation of the pulses. If the differential NPS of two optical beams cannot exceed this negative feedback range, the mode-locking can not be achieved. In the B $\rightarrow$ C process, the fiber loop shows positive feedback. Although the B $\rightarrow$ A and B $\rightarrow$ C processes are in opposite traces on the horizontal axis, they both show saturable absorption.
 
Since the slope at point B is zero, the transmittance differences for low-intensity lights around point B are close to zero and do not offer any intensity-dependent transmittance. In order to weaken the negative impacts of the process A $\rightarrow$ B and point B on differential NPS accumulation, a simple method is adding a $\pi/2$ phase shifter to the fiber loop, so as to move points A, B, and C to points A$'$, B$'$, and C$'$, respectively. As a result, both the initial transmittance of the light and the slope of the initial point will increase. When the phase shifter is inserted, the trace of the differential NPS is still B$'$ $\rightarrow$ A$'$ $\rightarrow$ B$'$ $\rightarrow$ C$'$, but the feedback type has changed. In the B$'$ $\rightarrow$ A$'$ process, the fiber loop shows negative feedback. In the A$'$ $\rightarrow$ B$'$ $\rightarrow$ C$'$ process, the fiber loop shows positive feedback. The phase shifter prevents the light in the fiber loop from forming pulses in the opposite traces such as B $\rightarrow$ A and B $\rightarrow$ C. It reduces the difficulty of turning around at point A and helps to accumulate the differential NPS in the correct trace when the sign change of differential NPS occurs.

After self-starting of mode locking, the pump power can be gradually reduced to a low level, while mode-locking still keeps. The mode-locking stability should be related to the fact that the pulsed light in the NALM does not experience the A $\rightarrow$ B negative feedback region. When the pump power is too low to support mode locking, the light will change from pulsed light to continuous light, and the output optical power will suddenly increase\cite{Grudinin1992}. This can be attributed to the sign change of differential NPS mentioned above.

In addition to increasing the power pump and introducing a phase shifter, the mode locking can also be achieved by adding a polarization controller on the fiber loop, tapping the fiber, slightly rotating the wave plate in the phase shifter, using an intra-cavity amplitude modulator to initiate the pulses\cite{Nicholson2006,Baumann2009}, or putting a moving mirror at the output port\cite{Wu1993}. Besides, it has been verified in our experiment that mode locking of a figure-9 fiber laser with the configuration in Ref.~\cite{Hansel2017} can be achieved by placing a moving aluminum alloy plate or a piece of paper at one of the two PBS output ports after the pump light is injected. While these methods can reduce the mode-locking threshold, one disadvantage is that after actively starting the mode-locking, the mode-locking cannot self-start after restarting the laser. Another disadvantage is that the same characteristics of the output pulse are also difficult to reproduce. For example, the pulse width may vary widely. The main principle of actively start mode-locking is that these methods suddenly change the phase of the laser, which is equivalent to directly inputting high-power optical signal to skip the negative feedback region of NALM or NOLM, rather than gradually cyclically amplifying the small optical signal in the cavity to evolve a stable phase shift. Moreover, in the experiment, we found that placing a static aluminum alloy plate or a piece of static paper at one of the two PBS output ports before the pump light is injected can make the self-starting threshold of the laser’s mode-locking higher or prevent self-starting.
\subsection{Main factors affecting NALM loop asymmetry}
Apart from ISA, the difficulty of mode locking self-starting is also related to the transmittance for the initial state, the slope of the transmittance curve, and the slope sign\cite{Hansel2017,Honda2017,Liao2019}, which are also closely related to the loop asymmetry of NALM. Gain-fiber position, splitting ratio, and optical attenuator in fiber loop are the three key factors that influence loop asymmetry.
\subsubsection{Gain-fiber position}
\begin{figure}[htbp]
	\centering\includegraphics[width=0.95\textwidth]{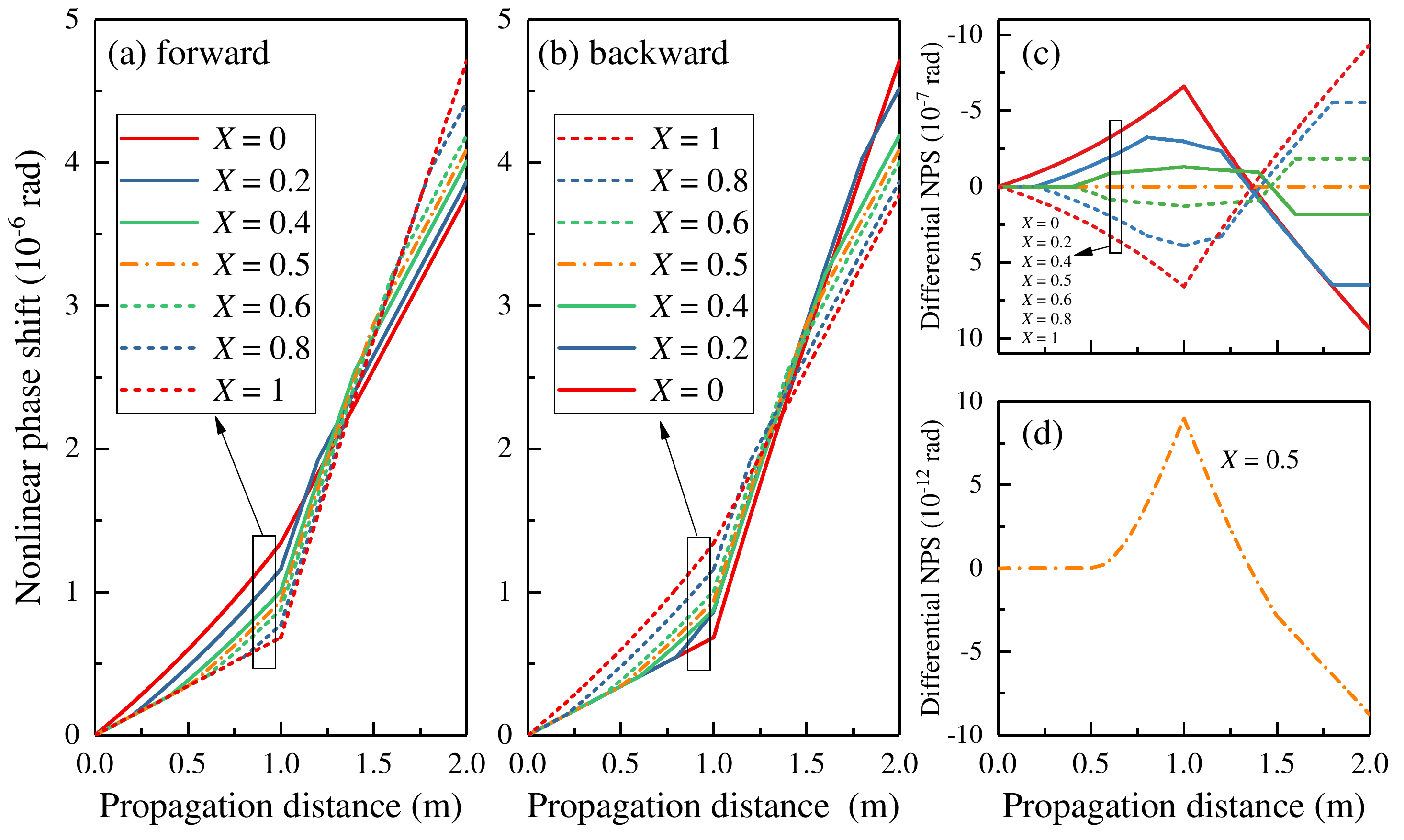}
	\caption{(a) The NPS of the forward beam as a function of propagation distance at different $X$. (b) The NPS of the backward beam. (c) The differential NPS between the forward and backward beams. (d) The differential NPS between the forward and backward beams at $X= 0.5$ m. In the calculation model, the erbium-doped fiber length $Y = 1$ m, the passive fiber length $X + Z = 1$ m, and the splitting ratio $\rho$ is 0.5.}
	\label{fig3}
\end{figure}

The NPS changes of the two small CW signals in passive fiber and EDFA are calculated to study the effect of gain-fiber position on NALM loop asymmetry. The results are given in Fig.~\ref{fig3}, which shows the differential NPS accumulation of forward and backward beams in the fiber loop with the gain fiber at different positions. When the two counterpropagating beams meet at $X = 1$ m, the speeds of the forward and backward NPS accumulations begin to change significantly, that is, the curve with the slower accumulation speed becomes the faster. This causes differential NPSs in Fig.~\ref{fig3}(c) to approach zero and the changes of differential NPS signs. When the gain fiber is placed in the symmetrical position ($X = 0.5$ m), the round-trip differential NPS shown in Fig.~\ref{fig3}(d) is negative because of the gain difference between forward and backward pump. When the gain fiber is not there, the farther the gain fiber is from port N, the larger the round-trip NPS of the forward beam is, and the smaller the round-trip NPS of the backward beam is. This means that the round-trip NPS of the CW light entering the EDFA earlier is smaller, which is not in line with the experience that the NPS of the earlier amplified pulsed light in the fiber loop is larger than that of the later amplified when only considering SPM.

It can be easily concluded from Fig.~\ref{fig3} that when the gain fiber is connected to one of the coupler ports that make up the fiber loop, the loop asymmetry is the largest. But this conclusion is not usually used in experiments. It is also necessary to consider the influence of WDM, which is not only related to the pump direction and the range of the pump effect gain fiber, but also often makes the pump gain fiber not directly connected to the port N or M in Fig.~\ref{fig1}. In 2011, A. Wang et al. designed the integrated WDM-collimator to solve this problem\cite{Wang2011}. Based on this, they then used all-gain-fiber loop to solve the problem of small gain caused by short fiber\cite{Liu2018h}. At this point, the loop asymmetry cannot be optimized by adjusting the gain fiber position. Because the loop asymmetry of all-gain-fiber design only relies on the gain difference between the forward and backward pump, which has little effect on differential NPS. Therefore, researchers often promote mode-locking self-starting by changing the beam splitting ratio or the power difference between two counterpropagating beams instead of adjusting the gain fiber position.

\subsubsection{Splitting ratio}
\begin{figure}[htbp]
	\centering\includegraphics[width=0.95\textwidth]{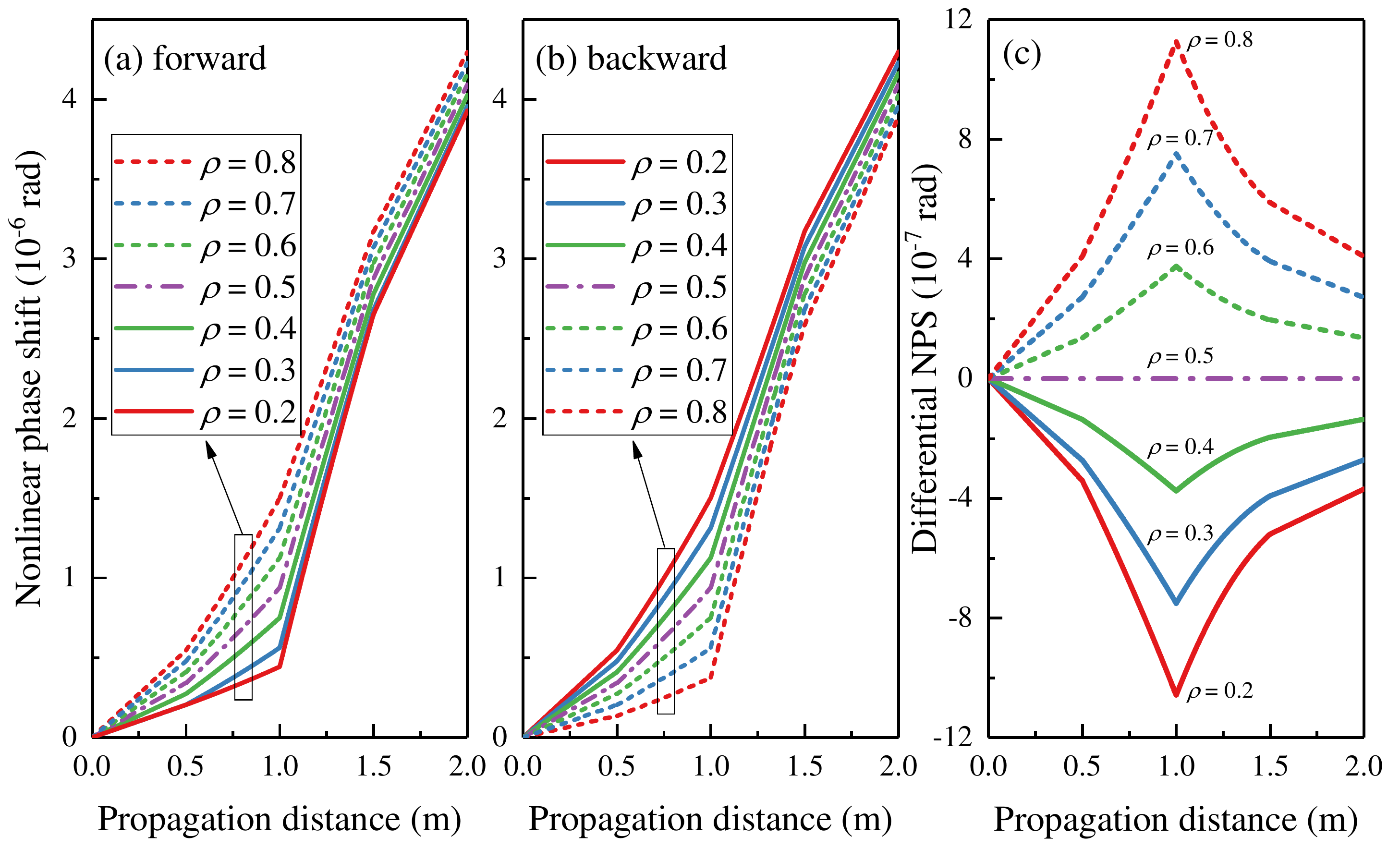}
	\caption{(a) The NPS of the forward beam as a function of the propagation distance under different splitting ratios. (b) The NPS of the backward beam. (c) The differential NPS between the forward and backward beams. In the calculation, the erbium-doped fiber length $Y = 1$ m, the passive fiber length $X = Z = 0.5$ m.}
	\label{fig4}
\end{figure}

 The nonlinear accumulation processes of the pulsed light in the NALM are calculated and given in Fig.~\ref{fig4} to discuss the effects of the splitting ratio on the self-starting of mode-locked fiber lasers with a NALM. The SPM phase shift of the pulsed light is proportional to its peak power, but for the XPM phase shift brought by the counterpropagating pulsed light in the fiber loop, the pulsed light experiences a pulse sequence with a uniform time and space interval. The peak-to-average power ratio of the pulse is a key factor in determining whether XPM dominates differential NPS\cite{Jinno1992,Clausen1996,Bogoni2004,Pitois2005}. Considering that the NPS is linear with optical power, the XPM phase shift can be calculated by the average power of the counterpropagating pulsed light. Because the ratios of peak power to average power of general mode-locked fiber lasers are in the order of $10^5$, the NPS of pulsed light is mainly determined by SPM effect as in Eq.~(\ref{eq:f-b_pulsed}).

Figure~\ref{fig4} shows the effects of splitting ratios on the NPS accumulations of forward and backward lights. Unlike Fig.~\ref{fig3}, there is no intersection of the curves. It is clear that the further the splitting ratio is away from 0.5, the more NPS accumulated. From the values of the differential NPS, for the forward light and the backward light with the same proportion of light intensity, the latter differential NPS is slightly larger than the former. But when the light intensity difference of split light is too large ($\rho = 0.2$ and $\rho = 0.8$), that is reversed.

\begin{figure}[htbp]
	\centering\includegraphics[width=0.6\textwidth]{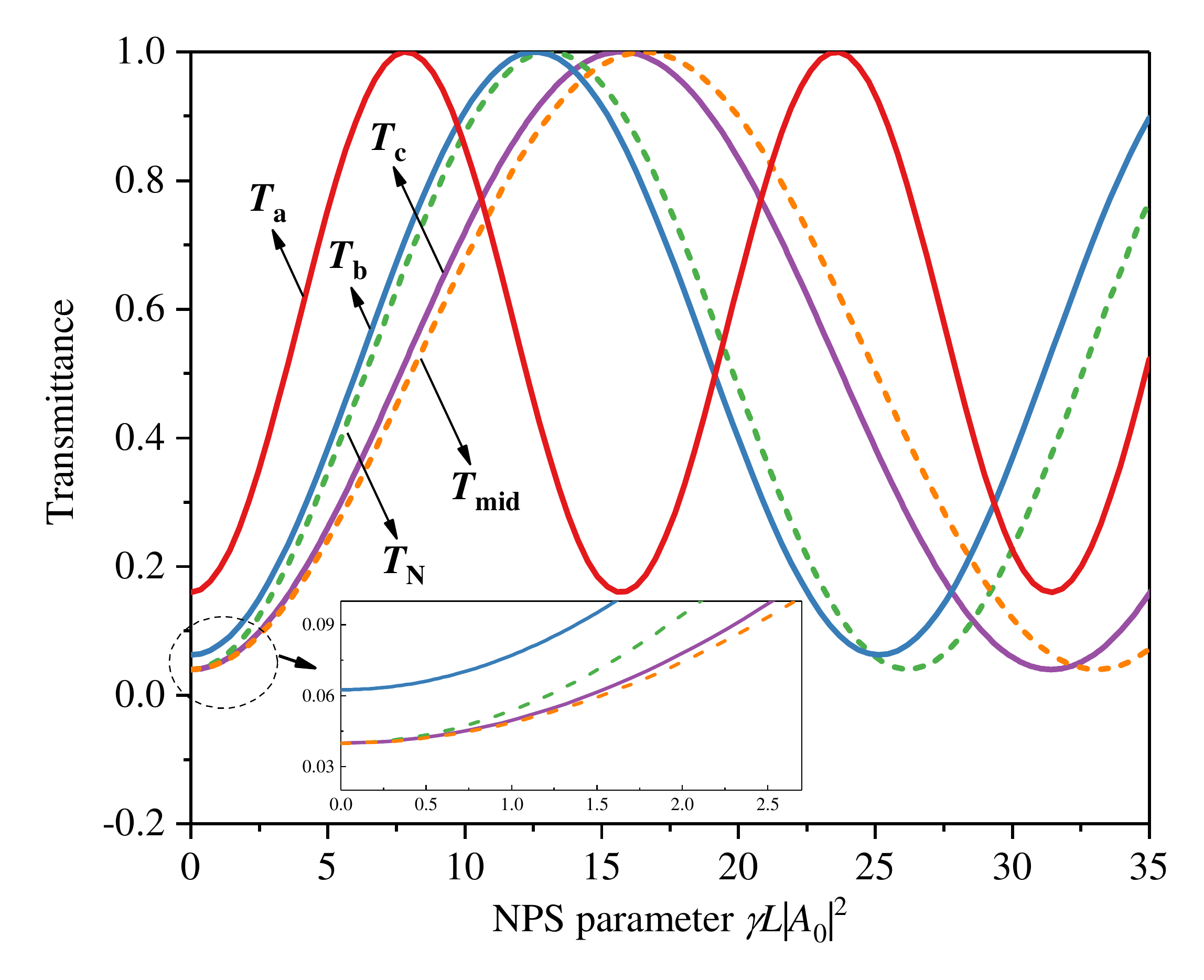}
	\caption{Variation of NOLM transmittance with NPS parameter $\gamma L|A_0|^2$. The $L$ is the length of the fiber loop. The five transmittance curves are plotted at different splitting ratios ($\rho$) and losses ($1-\alpha$), where $T_\mathrm{a}$: $\rho = 0.3$, $T_\mathrm{b}$: $\rho = 0.375$, $T_\mathrm{c}$: $\rho = 0.4$, $T_\mathrm{N}$: $\rho = 0.4, \alpha = 0.9$ and loss is added at port N, $T_\mathrm{mid}$: $\rho = 0.4, \alpha = 0.9$ and loss is added at the midpoint of the fiber between port N and port M.}
	\label{fig5}
\end{figure}

For simplicity, we employ the transmittance curve of the NOLM given in Fig.~\ref{fig5} to discuss the influence of the splitting ratio on the mode-locking and the saturable absorption effect related to the interference of the two beams in fiber loop. After a beam enters the fiber coupler, it is split into two counterpropagating parts. If the intensity difference between these two parts is too large, although the differential NPS required to go from the lowest point of the curve to the highest point will decrease, the amplitude related to the NOLM modulation depth of the curve will also decrease because the intensity difference is so large that their coherent interference is not obvious. In addition, the minimum transmittance of the NOLM is a parabolic function given by $T_{\rm{min}} = 1-\rho+2\rho^2$. As the splitting ratio $\rho$ is gradually away from 0.5, $T_{\rm{min}}$ at the zero-point increases, that is, the intracavity loss decreases, which is beneficial to the mode-locking self-starting. After considering these two factors, a fiber coupler with a splitting ratio of 0.3 to 0.4 is generally used to make a mode-locked fiber laser with a NALM.

Although both the splitting ratio and the optical attenuator can change the intensity ratio of the two counterpropagating beams, they have different effects on the round-trip transmittance of the NOLM. Figure~\ref{fig5} shows the influence of splitting ratio and optical attenuator on the transmittance of NOLM. By comparing the NPS parameters corresponding to the first maximum value of the three curves $T_\mathrm{a}$, $T_\mathrm{b}$, and $T_\mathrm{c}$, it can be seen that the closer the splitting ratio $\rho$ is to 0.5, the larger the NPS parameter is required to achieve the maximum transmittance. In addition, the minimum transmittance of the curve $T_\mathrm{c}$ is smaller than that of the curves $T_\mathrm{a}$ and $T_\mathrm{c}$, that is, the modulation depth of the NOLM corresponding to curve $T_\mathrm{c}$ is the largest.
\subsubsection{Optical attenuator}
To make self-starting easy, introducing losses in the optical fiber mirror is also a common method\cite{Bulushev1991,Seong2007,Wang2016a,Zhao2020d}. The losses mainly include fiber splice loss, laser output loss, and fiber device loss. The splice loss between two single-mode fibers assuming Gaussian mode shapes according to Marcuse's equation\cite{Marcuse1977} is given by
\begin{equation}
	\mathrm{Loss} \left( \mathrm{dB} \right) =-10\log \left\{ \frac{4D_{1}^{2}D_{2}^{2}}{\left( D_{1}^{2}+D_{2}^{2} \right) ^2}\exp \left[ -\frac{k_{0}^{2}n^2D_{1}^{2}D_{2}^{2}\sin ^2\theta}{2\left( D_{1}^{2}+D_{2}^{2} \right)} \right] \right\},
\end{equation}
where $D_1$ and $D_2$ are the mode field diameters (MFDs) of the two fibers, respectively. $\theta$ is the angular offset between the cores, $k_0$ is the vacuum wave number, and $n$ is the refractive index of the fibers’ cores. The splicing loss between the same fiber is generally around 0.02 dB. The main splicing loss comes from splicing between the passive and gain fibers. When the polarization-maintaining passive fiber (PM1550-XP) and the polarization-maintaining erbium-doped fiber (Er80-4/125-HD-PM, LIEKKI) are spliced without tapering, because the MFD of the erbium-doped fiber is smaller than that of passive fiber, the splicing loss is up to 0.82 dB. Suppose the fibers forming the optical fiber mirror are all erbium-doped. In that case, high energy density due to fiber gain and small MFD, as well as small splicing loss, can make the NPS accumulation easy and help the mode-locked fiber laser to achieve self-starting mode-locking\cite{Liu2018h}.

Losses in the fiber loop are not always harmful to self-starting. For simplicity, we ignore the gain fiber in Fig.~\ref{fig1} and only discuss NOLM next. Considering the phase shift $\pi/2$ brought by the coupler, the round-trip transmittance of NOLM is
\begin{equation}
	T\equiv \frac{\left| A_t \right|^2}{\left| A_0 \right|^2}=1-2\rho \left( 1-\rho \right) \left[ 1+\cos \left( \phi _f-\phi _b \right) \right].
	\label{eq:TTT}
\end{equation}
Here, the differential NPS is
\begin{equation}
	\phi _f-\phi _b=\left( 1-2\rho \right) \gamma \left| A_0 \right|^2L.
	\label{eq:F-B}
\end{equation}
When loss $1-\alpha$ is added to the coupler's forward light output port N, the transmittance becomes
\begin{equation}
	T_{\mathrm{N}}\equiv \frac{\left| A_t \right|^2}{\alpha \left| A_0 \right|^2}=1-2\rho \left( 1-\rho \right) \left\{ 1+\cos \left[ \left( 1-\rho -\alpha \rho \right) \gamma \left| A_0 \right|^2L \right] \right\},
	\label{eq:T_N}
\end{equation}
where the differential NPS is $(1-\rho -\alpha \rho) \gamma \left| A_0 \right|^2L$. If the light intensity of the forward light is greater than that of the backward light ($\rho > 0.5$), the absolute value of the differential NPS in Eq.~(\ref{eq:T_N}) is always smaller than that in Eq.~(\ref{eq:F-B}), that is, the loss is not helpful to the differential NPS accumulation. If $\rho < 0.5$, the loss is helpful, and it can increase the differential NPS and help lower self-starting threshold of the mode-locked laser with this NOLM. When the loss $1-\alpha$ is added at the intermediate position of the fiber between port M and port N, the transmittance becomes
\begin{equation}
	T_{\mathrm{mid}}\equiv \frac{\left| A_t \right|^2}{\alpha \left| A_0 \right|^2}=1-2\rho \left( 1-\rho \right) \left\{ 1+\cos \left[ \left( \alpha +1 \right) \left( \frac{1}{2}-\rho \right) \gamma \left| A_0 \right|^2L \right] \right\},
	\label{eq:T_mid}
\end{equation}
where the differential NPS is $(\alpha +1)(1/2 -\rho) \gamma \left| A_0 \right|^2L$, which is always smaller than that of Eq.~(\ref{eq:F-B}).

Under the same splitting ratio as curve $T_\mathrm{c}$, the curves $T_\mathrm{N}$ and $T_\mathrm{mid}$ in Fig.~\ref{fig5} are obtained after introducing a 10\% loss into the fiber loop. It can be seen that their highest points are located on either side of the highest point of curve $T_\mathrm{c}$. Unlike the case with different splitting ratios, the minimum transmissivities of curves $T_\mathrm{N}$ and $T_\mathrm{mid}$ are the same as that of curve $T_\mathrm{c}$, and the loss only changes the position of the maximum transmittance. To compare the different effects of splitting ratio and optical attenuator on NOLM more intuitively, we can consider the curves $T_\mathrm{N}$ and $T_\mathrm{b}$. The splitting ratio of curve $T_\mathrm{b}$ is set according to
\begin{equation}
	\rho'=\frac{\rho _c\left( 1-\alpha \right)}{1-\rho _c}
\end{equation}
where $\rho _c$ is the beam splitting ratio of the curve $T_\mathrm{c}$, $1-\alpha$ is the loss, and the $\alpha = 0.9$. Thus, the light intensities of two beams with a ratio of $\rho '$ are close to that of the two beams with a ratio of $\rho _c$ and a loss of $1-\alpha$ are added at the point N. It is clear that the curve $T_\mathrm{N}$ is only slightly shifted to the right relative to the curve $T_\mathrm{b}$, while the minimum transmittance is smaller than that of the curve $T_\mathrm{b}$.

We know that the closer $\rho$ is to 0.5, the greater the modulation depth of the NOLM. But when $\rho = 0.5$, NOLM does not show a saturation absorption effect. If the optical attenuator is added in NOLM, this problem can be avoided and the maximum modulation depth can be obtained. The optical attenuator here acts similarly to the gain fiber in NALM\cite{Zhao2020d}, but has no drawbacks: the gain of fiber is distributed along a line rather than at a point and the differential NPS between the two counterpropagating beams is not obvious when the gain fiber accounts for a large proportion of the fiber loop. Besides, before the two counterpropagating beams meet and interfere in the coupler, this loss can also effectively reduce the optical power difference introduced by the gain fiber and the coupler to make the NALM have a higher modulation depth.

Figure-9 fiber laser is one of the most popular mode-locked lasers at present. It has some characteristics different from those mentioned above. For the figure-9 fiber laser with a coupler, according to Eq.~(\ref{eq:TTT}), its transmittance is
\begin{equation}
	T'=1-T.
\end{equation}
Therefore, the transmittance curves are the reversal of the curves in Fig.~\ref{fig5}, and a phase shifter must be injected in the fiber loop so that the figure-9 configuration can show saturable absorption effect at the initial stage. The maximum transmittance is not equal to 1, and the minimum transmittance is zero. It can be seen that when the beam splitting ratio is far away from 0.5, the maximum transmittance will be reduced, and the large difference between the two interfering beams may raise the noise level intracavity\cite{Haus2000,Liao2019}, while controlling the energy ratio of the two beams through optical attenuator will avoid this situation. But for the figure-9 fiber laser with a PBS, the maximum and minimum transmittance will not change\cite{Duan2020}.

In practice, once the laser is built, it is often difficult to adjust the coupler's splitting ratio, while the loss can be continuously adjusted and compensated by increasing the gain in the cavity. The loss can also be induced via the output loss, where a suitable beam splitter can be freely introduced in the loop to improve loop asymmetry. The output loss of the fiber loop can be used for other applications, such as signal monitor and amplifier seed, and it is found in an experiment that the quality of the pulse output inside the loop is higher than that outside the loop\cite{Jiang2016}.
\section{Conclusion}
We study the self-starting of the mode-locked fiber laser with a NALM by calculating the phase shift evolution in the fiber loop. By comparing the round-trip differential NPS signs of the CW light and the pulses, we find that the NALM can behave as an inverse saturable absorber during the pulse formation process, which is not helpful for pulse formation and mode-locking self-starting. The impact of ISA can be alleviated by introducing a non-reciprocal phase shifter into the fiber loop and shifting the ISA region on the transmittance curve. The ISA could be used to explain the experimental phenomena that the mode-locking of fiber laser can be actively started by triggers such as tapping the fiber, fine-tuning the polarization controller or waveplate, and injecting noise. Besides, we analyze three factors that affect self-starting, including the gain-fiber position, the splitting ratio, and the optical attenuator in the fiber loop. It is found that although both the splitting ratio and the optical attenuator can change the intensity ratio of the two interfering beams, they have different effects on the round-trip transmittance curve of the optical loop mirror. The increase of beam splitting ratio can reduce the amplitude of transmittance curve and increase the slope of curve, while the optical attenuator does not change the amplitude of curve. At present, for the mode-locked fiber laser with a NALM, how to reduce the difficulty of manufacturing the high-repetition-rate fiber laser and how to lower the self-starting threshold are urgent problems to be solved. The discussion about the ISA mechanism and the three factors that affect self-starting are of potential help in these areas.

\section*{Funding}
National Natural Science Foundation of China (11604385, 62105368 and 91536106); Natural Science Foundation of Hunan Province, China (2019JJ50743); Research Project of National University of Defense Technology (ZK17-03-11).

\section*{Acknowledgments}
We would like to thank Dr. Ke Yin for the instruction on our work, Dr. Guochao Wang for the discussions about mode-locked fiber laser, and Joona Rissanen for the rate equation simulation library \href{https://github.com/Jomiri/pyfiberamp}{\texttt{PyFiberAmp}}.

\section*{Disclosures}
The authors declare no conflicts of interest.


\bibliography{Optica-template}

\begin{thebibliography}{10}
\newcommand{\enquote}[1]{``#1''}

\bibitem{Fermann1990}
M.~E. Fermann, F.~Haberl, M.~Hofer, and H.~Hochreiter, \enquote{Nonlinear
  amplifying loop mirror,} {\protect\JournalTitle{Opt. Lett.}} \textbf{15}, 752
  (1990).

\bibitem{Doran1988}
N.~J. Doran and D.~Wood, \enquote{Nonlinear-optical loop mirror,}
  {\protect\JournalTitle{Opt. Lett.}} \textbf{13}, 56 (1988).

\bibitem{Duling1991a}
I.~Duling, \enquote{Subpicosecond all-fibre erbium laser,}
  {\protect\JournalTitle{Electron. Lett.}} \textbf{27}, 544 (1991).

\bibitem{Fermann1991}
M.~E. Fermann, M.~Hofer, F.~Haberl, A.~J. Schmidt, and L.~Turi,
  \enquote{Additive-pulse-compression mode locking of a neodymium fiber laser,}
  {\protect\JournalTitle{Opt. Lett.}} \textbf{16}, 244--246 (1991).

\bibitem{Richardson1991}
D.~Richardson, R.~Laming, D.~Payne, V.~Matsas, and M.~Phillips,
  \enquote{Selfstarting, passively modelocked erbium fibre ring laser based on
  the amplifying {S}agnac switch,} {\protect\JournalTitle{Electron. Lett.}}
  \textbf{27}, 542 (1991).

\bibitem{Bulushev1991}
A.~G. Bulushev, E.~M. Dianov, and O.~G. Okhotnikov, \enquote{Self-starting
  mode-locked laser with a nonlinear ring resonator,}
  {\protect\JournalTitle{Opt. Lett.}} \textbf{16}, 88--90 (1991).

\bibitem{Seong2007}
N.~H. Seong, D.~Y. Kim, and S.~P. Veetil, \enquote{Mode-locked fiber laser
  based on an attenuation-imbalanced nonlinear optical loop mirror,}
  {\protect\JournalTitle{Optics Communications}} \textbf{280}, 438--442 (2007).

\bibitem{Zhao2020d}
J.~Zhao, J.~Zhou, Y.~Jiang, L.~Li, D.~Shen, A.~Komarov, L.~Su, D.~Tang,
  M.~Klimczak, and L.~Zhao, \enquote{Nonlinear absorbing-loop mirror in a
  holmium-doped fiber laser,} {\protect\JournalTitle{Journal of Lightwave
  Technology}} \textbf{38}, 6069--6075 (2020).

\bibitem{Bogoni2004}
A.~Bogoni, M.~Scaffardi, P.~Ghelfi, and L.~Poti, \enquote{Nonlinear optical
  loop mirrors: Investigation solution and experimental validation for
  undesirable counterpropagating effects in all-optical signal processing,}
  {\protect\JournalTitle{IEEE Journal of Selected Topics in Quantum
  Electronics}} \textbf{10}, 1115--1123 (2004).

\bibitem{Pitois2005}
S.~Pitois, \enquote{Influence of cross-phase modulation in {SPM}-based
  nonlinear optical loop mirror,} {\protect\JournalTitle{Optics
  Communications}} \textbf{253}, 332--337 (2005).

\bibitem{Ai2011}
F.~Ai, Z.~Cao, X.~Zhang, C.~Zhang, B.~Zhang, and B.~Yu, \enquote{Passively
  mode-locked fiber laser with kilohertz magnitude repetition rate and tunable
  pulse width,} {\protect\JournalTitle{Optics \& Laser Technology}}
  \textbf{43}, 501--505 (2011).

\bibitem{Steele1993}
A.~Steele, \enquote{Pulse compression by an optical fibre loop mirror
  constructed from two different fibres,} {\protect\JournalTitle{Electron.
  Lett.}} \textbf{29}, 1972 (1993).

\bibitem{Wong1997a}
W.~S. Wong, S.~Namiki, M.~Margalit, H.~A. Haus, and E.~P. Ippen,
  \enquote{Self-switching of optical pulses in dispersion-imbalanced nonlinear
  loop mirrors,} {\protect\JournalTitle{Opt. Lett.}} \textbf{22}, 1150 (1997).

\bibitem{Matsumoto1998}
M.~Matsumoto and T.~Ohishi, \enquote{Dispersion-imbalanced nonlinear optical
  loop mirror with lumped dispersive elements,}
  {\protect\JournalTitle{Electron. Lett.}} \textbf{34}, 1140 (1998).

\bibitem{Yongkui2003}
D.~Yongkui, W.~Yongqiang, L.~Zhiyong, T.~Li, and L.~Shichen, \enquote{Pulse
  amplitude equalization in a harmonically modelocked fiber laser using a
  dispersion imbalanced non-linear loop mirror,} {\protect\JournalTitle{Optics
  Communications}} \textbf{225}, 363--369 (2003).

\bibitem{Mortimore1988}
D.~Mortimore, \enquote{Fiber loop reflectors,} {\protect\JournalTitle{Journal
  of Lightwave Technology}} \textbf{6}, 1217--1224 (1988).

\bibitem{Steele1996}
A.~L. Steele and J.~P. Hemingway, \enquote{Nonlinear optical loop mirror
  constructed from dispersion decreasing fibre,} {\protect\JournalTitle{Optics
  Communications}} \textbf{123}, 487--491 (1996).

\bibitem{Shah2001}
D.~D. Shah, J.~Ravikanth, and R.~Vijaya, \enquote{Optimization of peak
  transmittivity in a non-linear fiber loop mirror\textemdash an accurate
  analysis,} {\protect\JournalTitle{Optics Communications}} \textbf{197},
  301--308 (2001).

\bibitem{Meissner2005}
M.~Meissner, M.~R{\"o}sch, B.~Schmauss, and G.~Leuchs, \enquote{Optimum
  splitting ratio for amplifier noise reduction by an asymmetric nonlinear
  optical loop mirror,} {\protect\JournalTitle{Appl. Phys. B}} \textbf{80},
  489--495 (2005).

\bibitem{Kane1994}
M.~G. Kane, I.~Glesk, J.~P. Sokoloff, and P.~R. Prucnal, \enquote{Asymmetric
  optical loop mirror: Analysis of an all-optical switch,}
  {\protect\JournalTitle{Appl. Opt.}} \textbf{33}, 6833 (1994).

\bibitem{Sponsel2007}
K.~Sponsel, K.~Cvecek, C.~Stephan, G.~Onishchukov, B.~Schmauss, and G.~Leuchs,
  \enquote{Optimization of a nonlinear amplifying loop mirror for amplitude
  regeneration in phase-shift-keyed transmission,} {\protect\JournalTitle{IEEE
  Photonics Technology Letters}} \textbf{19}, 1858--1860 (2007).

\bibitem{Hierold2011}
M.~Hierold, T.~Roethlingshoefer, K.~Sponsel, G.~Onishchukov, B.~Schmauss, and
  G.~Leuchs, \enquote{Multilevel phase-preserving amplitude regeneration using
  a single nonlinear amplifying loop mirror,} {\protect\JournalTitle{IEEE
  Photon. Technol. Lett.}} \textbf{23}, 1007--1009 (2011).

\bibitem{Zhang2021c}
S.~Zhang, B.~Wu, and F.~Wen, \enquote{Phase-preserving {NALM} regenerator with
  lower input power by optimizing the nonreciprocal phase shifter,}
  {\protect\JournalTitle{Appl. Opt.}} \textbf{60}, 492 (2021).

\bibitem{HongLin1994}
{Hong Lin}, D.~Donald, and W.~Sorin, \enquote{Optimizing polarization states in
  a figure-8 laser using a nonreciprocal phase shifter,}
  {\protect\JournalTitle{J. Lightwave Technol.}} \textbf{12}, 1121--1128
  (1994).

\bibitem{Moores1991}
J.~D. Moores, K.~Bergman, H.~A. Haus, and E.~P. Ippen, \enquote{Optical
  switching using fiber ring reflectors,} {\protect\JournalTitle{J. Opt. Soc.
  Am. B}} \textbf{8}, 594 (1991).

\bibitem{Finlayson1992}
N.~Finlayson, B.~K. Nayar, and N.~J. Doran, \enquote{Switch inversion and
  polarization sensitivity of the nonlinear-optical loop mirror,}
  {\protect\JournalTitle{Opt. Lett.}} \textbf{17}, 112 (1992).

\bibitem{HeeYealRhy2000}
{Hee Yeal Rhy}, {Byoung Yoon Kim}, and {Hai-Woong Lee}, \enquote{Self-switching
  with a nonlinear birefringent loop mirror,} {\protect\JournalTitle{IEEE J.
  Quantum Electron.}} \textbf{36}, 89--93 (2000).

\bibitem{Pottiez2004a}
O.~Pottiez, E.~A. Kuzin, B.~{Ibarra-Escamilla}, J.~T. {Camas-Anzueto}, and
  F.~{Guti{\'e}rrez-Zainos}, \enquote{Easily tunable nonlinear optical loop
  mirror based on polarization asymmetry,} {\protect\JournalTitle{Opt.
  Express}} \textbf{12}, 3878 (2004).

\bibitem{WanliGao2018}
W.~Gao, G.~Liu, and Z.~Zhang, \enquote{44.6 fs pulses from a 257 {MH}z
  {Er}:fiber laser mode-locked by biased {NALM},} {\protect\JournalTitle{Chin.
  Opt. Lett.}} \textbf{16}, 111401 (2018).

\bibitem{Hansel2017}
W.~H{\"a}nsel, H.~Hoogland, M.~Giunta, S.~Schmid, T.~Steinmetz, R.~Doubek,
  P.~Mayer, S.~Dobner, C.~Cleff, M.~Fischer, and R.~Holzwarth, \enquote{All
  polarization-maintaining fiber laser architecture for robust femtosecond
  pulse generation,} {\protect\JournalTitle{Appl. Phys. B}} \textbf{123}, 41
  (2017).

\bibitem{Jinno1992}
M.~Jinno and T.~Matsumoto, \enquote{Nonlinear {S}agnac interferometer switch
  and its applications,} {\protect\JournalTitle{IEEE Journal of Quantum
  Electronics}} \textbf{28}, 875--882 (1992).

\bibitem{Clausen1996}
C.~B. Clausen, J.~H. Povlsen, and K.~Rottwitt, \enquote{Polarization
  sensitivity of the nonlinear amplifying loop mirror,}
  {\protect\JournalTitle{Opt. Lett.}} \textbf{21}, 1535 (1996).

\bibitem{Keller2021}
U.~Keller, \emph{Ultrafast Lasers: A Comprehensive Introduction to Fundamental
  Principles with Practical Applications} (Springer Nature Switzerland AG,
  Switzerland, 2021).

\bibitem{Malcuit1984}
M.~S. Malcuit, R.~W. Boyd, L.~W. Hillman, J.~Krasinski, and C.~R. Stroud,
  \enquote{Saturation and inverse-saturation absorption line shapes in
  alexandrite,} {\protect\JournalTitle{J. Opt. Soc. Am. B}} \textbf{1}, 73--75
  (1984).

\bibitem{Rakov2001}
N.~Rakov, C.~B. {de Ara{\'u}jo}, G.~B. Rocha, A.~M. Simas, P.~A.~F.
  {Athayde-Filho}, and J.~Miller, \enquote{Reverse saturable absorption and
  anti-{S}tokes fluorescence in mesoionic compounds pumped at 532 nm,}
  {\protect\JournalTitle{Appl. Opt.}} \textbf{40}, 1389 (2001).

\bibitem{Li2014}
X.~Li, Y.~Wang, Y.~Wang, W.~Zhao, X.~Yu, Z.~Sun, X.~Cheng, X.~Yu, Y.~Zhang, and
  Q.~J. Wang, \enquote{Nonlinear absorption of {SWNT} film and its effects to
  the operation state of pulsed fiber laser,} {\protect\JournalTitle{Opt.
  Express}} \textbf{22}, 17227--17235 (2014).

\bibitem{Cheng2015}
Z.~Cheng, H.~Li, H.~Shi, J.~Ren, Q.-H. Yang, and P.~Wang, \enquote{Dissipative
  soliton resonance and reverse saturable absorption in graphene oxide
  mode-locked all-normal-dispersion {Yb}-doped fiber laser,}
  {\protect\JournalTitle{Opt. Express}} \textbf{23}, 7000--7006 (2015).

\bibitem{Tian2018}
X.~Tian, R.~Wei, Q.~Guo, Y.-J. Zhao, and J.~Qiu, \enquote{Reverse saturable
  absorption induced by phonon-assisted anti-{S}tokes processes,}
  {\protect\JournalTitle{Adv. Mater.}} \textbf{30}, 1801638 (2018).

\bibitem{Giles1991}
C.~Giles and E.~Desurvire, \enquote{Modeling erbium-doped fiber amplifiers,}
  {\protect\JournalTitle{J. Lightwave Technol.}} \textbf{9}, 271--283 (1991).

\bibitem{Agrawal2019}
G.~Agrawal, \emph{Nonlinear Fiber Optics} (Elsevier Inc, Cambridge, 2019),
  sixth ed.

\bibitem{Duan2020}
D.~Duan, J.~Wang, Y.~Wu, J.~Ma, and Q.~Mao, \enquote{Approach to high pulse
  energy emission of the self-starting mode-locked figure-9 fiber laser,}
  {\protect\JournalTitle{Optics Express}} \textbf{28}, 33603--33613 (2020).

\bibitem{Grudinin1992}
A.~Grudinin, D.~Richardson, and D.~Payne, \enquote{Energy quantisation in
  figure eight fibre laser,} {\protect\JournalTitle{Electronics Letters}}
  \textbf{28}, 67--68 (1992).

\bibitem{Nicholson2006}
J.~W. Nicholson and M.~Andrejco, \enquote{A polarization maintaining,
  dispersion managed, femtosecond figure-eight fiber laser,}
  {\protect\JournalTitle{Opt. Express}} \textbf{14}, 8160 (2006).

\bibitem{Baumann2009}
E.~Baumann, F.~R. Giorgetta, J.~W. Nicholson, W.~C. Swann, I.~Coddington, and
  N.~R. Newbury, \enquote{High-performance, vibration-immune, fiber-laser
  frequency comb,} {\protect\JournalTitle{Opt. Lett.}} \textbf{34}, 638 (2009).

\bibitem{Wu1993}
S.~Wu, T.~F. Morse, J.~Strait, and R.~L. Fork, \enquote{High-power passively
  mode-locked {Er}-doped fiber laser with a nonlinear optical loop mirror,}
  {\protect\JournalTitle{Opt. Lett.}} \textbf{18}, 1444 (1993).

\bibitem{Honda2017}
T.~Honda, S.~Y. Set, and S.~Yamashita, \enquote{Effects of non-reciprocal phase
  bias in figure-8/9 fiber lasers,} in \emph{Conference on {Lasers} and
  {Electro}-{Optics},}  (OSA, San Jose, California, 2017), p. SM4L.7.

\bibitem{Liao2019}
R.~Liao, Y.~Song, L.~Chai, and M.~Hu, \enquote{Pulse dynamics manipulation by
  the phase bias in a nonlinear fiber amplifying loop mirror,}
  {\protect\JournalTitle{Opt. Express}} \textbf{27}, 14705 (2019).

\bibitem{Wang2011}
A.~Wang, H.~Yang, and Z.~Zhang, \enquote{503 {M}hz repetition rate femtosecond
  {Yb}:fiber ring laser with an integrated {WDM} collimator,}
  {\protect\JournalTitle{Optics Express}} \textbf{19}, 25412 (2011).

\bibitem{Liu2018h}
G.~Liu, X.~Jiang, A.~Wang, G.~Chang, F.~Kaertner, and Z.~Zhang, \enquote{Robust
  700 {MH}z mode-locked {Yb}:fiber laser with a biased nonlinear amplifying
  loop mirror,} {\protect\JournalTitle{Opt. Express}} \textbf{26}, 26003
  (2018).

\bibitem{Wang2016a}
W.~Wang, F.~Wang, Y.~Zhang, H.~Ma, Q.~Yu, and X.~Zhang, \enquote{Passively
  mode-locked figure-eight fiber laser using a compact power-imbalanced
  nonlinear optical-loop mirror,} {\protect\JournalTitle{Journal of Russian
  Laser Research}} \textbf{37}, 265--272 (2016).

\bibitem{Marcuse1977}
D.~Marcuse, \enquote{Loss analysis of single-mode fiber splices,}
  {\protect\JournalTitle{Bell System Technical Journal}} \textbf{56}, 703--718
  (1977).

\bibitem{Haus2000}
H.~Haus, \enquote{Mode-locking of lasers,} {\protect\JournalTitle{IEEE Journal
  of Selected Topics in Quantum Electronics}} \textbf{6}, 1173--1185 (2000).

\bibitem{Jiang2016}
T.~Jiang, Y.~Cui, P.~Lu, C.~Li, A.~Wang, and Z.~Zhang, \enquote{All {PM} fiber
  laser mode locked with a compact phase biased amplifier loop mirror,}
  {\protect\JournalTitle{IEEE Photon. Technol. Lett.}} \textbf{28}, 1786--1789
  (2016).

\end{thebibliography}

\end{document}